\documentstyle [11pt,graphicx] {article}
\input{epsf}
\begin{document}
\renewcommand{\baselinestretch}{1.5}
\begin{center}
\huge {\bf  Acoustic Oscillations
in the Early Universe and Today}
\end{center}

\normalsize

\vspace{1cm}

\begin{center}

{\bf Christopher J. Miller, Robert C. Nichol,\\ Department of Physics, \\ Carnegie Mellon University,
\\ Pittsburgh, PA 15213,\\ United States of America.\\}

\vspace{1cm}

 {\bf and}\\

\vspace{1cm}

{\bf David J. Batuski,\\ Department of Physics \& Astronomy, \\  University of Maine, \\
Orono, ME, 04469\\ United States of America.\\}

\end{center}

\vspace{1cm}
\noindent{\bf During its first $\simeq100,000$ years, the
universe was a fully ionized plasma with a tight coupling by Thompson scattering between the photons
and matter. 
The trade--off between gravitational collapse and photon pressure causes
acoustic oscillations in this primordial fluid. These oscillations will leave
predictable imprints
in the spectra of the cosmic microwave background and the present day matter-density
distribution.
Recently, the BOOMERANG and
MAXIMA teams announced the detection of these
acoustic oscillations in the cosmic microwave background (observed at redshift $ \simeq 1000$).
Here, we compare these CMB detections
with the corresponding acoustic oscillations in the matter-density
power spectrum (observed at redshift  $\simeq 0.1$).
These consistent results, from two different cosmological epochs,
provide further support for
our standard Hot Big Bang model
of the universe.}

The standard model of cosmology is the Inflationary Hot Big Bang
scenario. A key aspect of this model is the ease with which it
explains some critical observational facts about the universe.
For example, the existence of the cosmic microwave background
(CMB) radiation that fills all space is simply the radio remnant of
a hot early phase of the universe {\it i.e.} when it was only
$\simeq$ 100,000 years old. The model also provides a natural
explanation for Hubble's famous expansion, large--scale coherent
structures in the mass distribution (caused by quantum effects
in the early universe), as well as producing a flat global geometry
for the universe ({\it 1}).
In this scenario, the distribution
of matter on the largest scales is connected, through well-established 
physics, to the temperature fluctuations in the CMB. 
Thus, any independent agreement between the CMB (at redshift $\simeq 1000$) and
the matter-density distribution (at redshift $\simeq 0.1$), is naturally explained 
by the Hot Big Bang Inflationary model.

The early universe was a plasma made uo of photons, electrons
and protons, along with the so-called Dark Matter. 
During this period, the gravitational force from potential wells (created as a result of
local curvature pertubations or dark matter clumps) causes compressions in this fluid.
As the plasma collapses inward, it meets resistance
from photon pressure, reversing the plasma direction and causing a subsequent rarefaction.
This cycle of compression and rarefaction results in acoustic
oscillations, where baryons act as a source of
inertia.  Compression (rarefaction)
of the plasma creates hot (cold) spots in the temperature of the
plasma. 
Because
the photons and baryons are coupled through
Thompson scattering, the matter-density power spectrum will also exhibit these
oscillations.  
As the universe cooled and the photons and matter decoupled,
the acoustic oscillations became
frozen as oscillatory features in both the temperature and matter-density power spectra.
These acoustic oscillations are a general prediction from
gravitational instability models of structure formation ({\it 2,3}).

The recent results from the MAXIMA and BOOMERANG CMB balloon experiments 
provide evidence for the first two acoustic peaks ({\it 4--8}). These
acoustic oscillations are the peaks and valleys in {\bf Fig. 1A}.
The location and amplitude of the first peak indicates
that we live in a universe that is geometrically flat, with a
fractional contribution of the total mass over the critical mass (required for
a flat universe) of $\Omega_{matter}h^2 \simeq 0.2$, where the reduced Hubble
constant is $h = H_0/100$ km s$^{-1}$Mpc$^{-1}$. The remaining contribution to
the critical mass must come in some form of dark energy (specified as
$\Omega_{vacuum}$). 
The CMB data also indicate the detection of the second acoustic peak
that, along with the first peak, can constrain the density of baryons in the
universe ($\Omega_{baryons}$). 

Until recently, the use of the local matter distribution in the universe 
has been limited to constraining just $\Omega_{matter}h$,
as the
datasets were not large enough to detect the acoustic oscillations.
On scales smaller than $\sim 50h^{-1}$Mpc, the oscillations will be wiped out
by the individual motions of galaxies and clusters. 
Here, we examine the
matter-density power spectrum on near-gigaparsec scales, where the imprint of the acoustic
oscillations should be detectable.
For our study, we use clusters of galaxies and individual galaxies as tracers of
the matter in the universe, and
describe their distribution by the
power spectrum, $P(k)$ of fluctuations in this density field, $\delta({\bf r})$:
\begin{equation}
\delta({\bf r}) = \frac{\rho({\bf r}) - \langle \rho \rangle}{\langle \rho \rangle}.
\end{equation}
The power spectrum is a function of wavenumber $k = 2\pi/\lambda$, where
$\lambda$ is the scale size in units of $h^{-1}$Mpc.
We have derived $P(k)$ from three cosmological redshift
surveys: the Abell/ACO Cluster Survey ({\it 9,10}), the IRAS Point Source redshift
catalog (PSCz) ({\it 11,12}), and the Automated Plate Measuring machine (APM) cluster
catalog ({\it 13,14}).  The volumes traced by these surveys
are large enough to accurately probe the power spectrum to 
near-gigaparsec scales.

We find oscillatory features in the matter-density power spectrum
({\bf Fig 1B}), consistent with 
a cosmological model having 
$\Omega_{matter}h^2 = 0.12^{+0.02}_{-0.03}$,
$\Omega_{baryons}h^2 = 0.029^{+0.011}_{-0.015}$, and $n_s = 1.08^{+0.17}_{-0.20}$, where
$n_s$ is the primordial spectra index ($2\sigma$ confidence limits) {\it (15,19)}.
These fitted parameters
provide almost enough information 
to independently predict the CMB temperature spectrum free of any CMB data.
All that is needed is a choice for $\Omega_{vacuum}$,
which affects the temperature
power spectrum, but has no effect on the shape of the local matter-density
power spectrum {\it (16)}.
Fortunately, the recent Type Ia supernovae results provide
us with an independent measurement of $\Omega_{vacuum}$ {\it (17,18)}.
Thus, using the data at redshift $\simeq 0.1$
from galaxies and clusters of 
galaxies, along with the recent supernovae data at redshift $\simeq 1$,
we can accurately predict
the CMB temperature power spectrum (Fig. 1A) at redshift $\simeq 1000$, under
the assumption of the standard cosmological model.

We see a direct concordance between the CMB, which originated $\simeq$ 100,000 years
after the Big Bang, the supernovae data, measured at roughly half
the age of the universe,
and the matter-density distribution,  which is measured today.
Not only do these results provide support for
the Hot Big Bang Inflationary model, they also show that we
understand the physics of the early universe. This physics can
take us forward in time, predicting the matter-density distribution
from the CMB, or as we have shown here, backwards in time, ``predicting''
the CMB using the distribution of galaxies and clusters in our local
universe.

\newpage

\vspace{0.5in}

\begin{figure}[h]
\scalebox{0.8}{\includegraphics{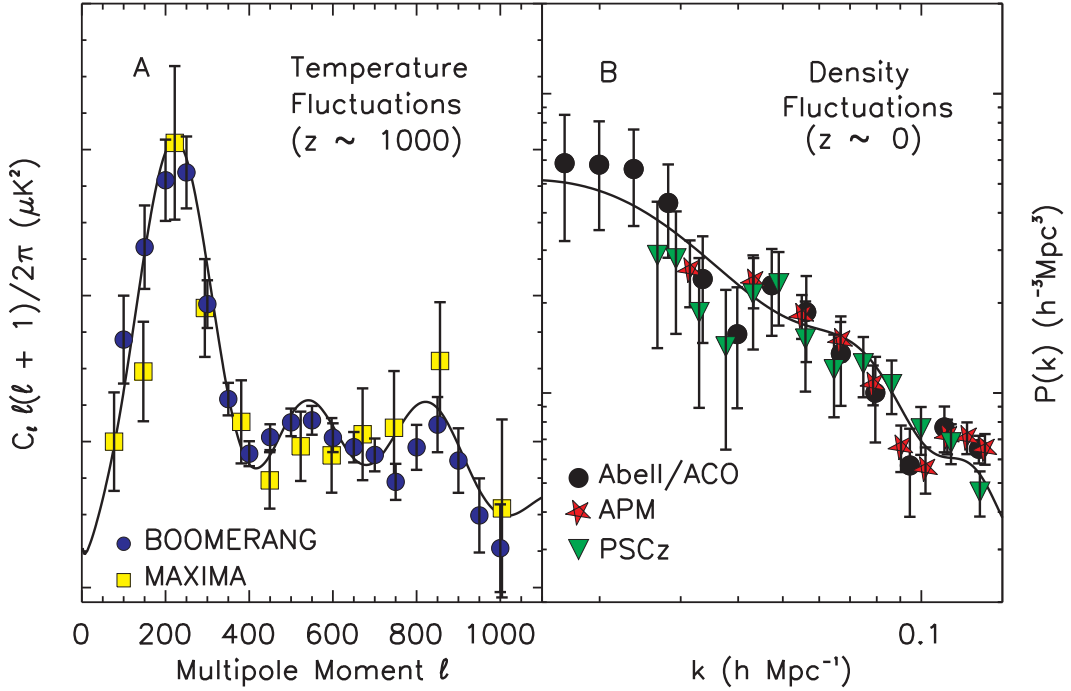}}
\caption[]{We plot the CMB data from the MAXIMA and BOOMERANG experiments
({\bf left}) along side the matter-density data ({\bf right}).
The solid line is the best fit model ($\Omega_m = 0.24$, $\Omega_b
= 0.06$, and $n_s = 1.08$ with $H_0 = 69$) using the matter-density data alone.
The amplitudes in both plots remain a free parameter.
The solid line in {\bf (A)}  is {\it not} a fit
to the CMB data (although the $\chi^2$ is 34 for 32 data points).
It is the resultant cosmological model
using the best-fit parameters from {\bf (B)} and
$\Omega_{vacuum}=0.8$, consistent with the Type Ia supernovae results {\it (17)}.
}
\end{figure}
\vspace{0.5in}
\pagebreak
\newpage

{\large {\bf References}}
 \everypar=
  {\hangafter=1 \hangindent=.4in
    \baselineskip=\singlespace}

\begin{tabbing}
Born\= on a mountain top in tennessee! \kill

\noindent 1. Peebles, P.\ J.\ E.\, {\em Principles of Physics Cosmology}
(Princeton University Press, NJ,\\
\>1993).\\
\noindent 2. Peebles, P.\ J.\ E.\ \& 
Yu, J.\ T.,\
{\em  Astrophys. J.} {\bf 162}, 815 (1970).\\
\noindent 3. Hu, W.\ \& White, M.,\ 
{\em  Astrophys. J.} {\bf 471}, 30 (1996).\\
\noindent 4. Miller, A.D., et al. {\em Astrophys. J.} {\bf 524L}, 1 (1999).\\
\noindent 5. Balbi, A.\ et al.,\ 
{\em  Astrophys. J.} {\bf 545}, L1 (2000). \\ 
\noindent 6.  Melchiorri, A.\ et
al., {\em  Astrophys. J.} {\bf 536}, L63 (2000).\\
\noindent 7. Netterfield, C.B., et al. submitted to {\em Astrophys. J.} (2001) astro-ph/0104460. \\
\noindent 8. Lee, A.T., et al. (2001) astro-ph/0104459 \\
\noindent 9. Miller, C.J., Batuski, D.J., {\em  Astrophys. J.} in press (2001). \\
\noindent 10. Miller, C.J., Krughoff, K.S.,  Batuski, D.J., Slinglend, K.A., Hill, J.M., \\
\> submitted to {\em Astron. J.} (2001).\\
\noindent 11. Saunders, W.\ et al.,\
{\em Mon. Not. R. astr. Soc.} {\bf 317}, 55 (2000).\\
\noindent 12. Hamilton, A.J.S., Tegmark, M.,  submitted to {\em Mon. Not. R. astr. Soc.} (2001)\\
\> astro-ph/0008392.\\
\noindent 13. Dalton, G.B.,
Efstathiou, G., Maddox, S.J., and Sutherland, W.J., \\
\> {\em Mon. Not. R. astr. Soc.} 
 {\bf 271}, 47 (1994).\\
\noindent 14. Tadros, H.,Efstathiou, G., \& Dalton, G.,
  {\em Mon. Not. R. astr. Soc.} {\bf 296}, 995 (1998).\\
\noindent 15. Miller, C.J., Nichol, R.C. \& Batuski, D.J.,
{\em  Astrophys. J.} in press (2001).\\
\noindent 16. This is only true if we fix the Hubble constant,
$\Omega_{matter}$, and  $\Omega_{vacuum}$, \\
\>thus altering only the curvature of the universe, which has no effect on the shape of \\
\>the matter-density
power spectrum.\\
\noindent 17. Perlmutter, S.\ et
al.,
{\em Astrophys. J.} 517, 565 (1999).\\
\noindent 18.  Reiss, A.G, Press, W.H., Kirshner, R.P. {\em Astrophys. J.}
{\bf 438L}, 17 (1995)\\
\noindent 19. We use the publicly available CMBFAST code to generate our cosmological models. \\
\>  This code is described in Seljak, U. \& Zaladarriaga, M., {\em Astrophys. J.} {\bf 469},\\
\>  437 (1996) \\
\end{tabbing}

\end{document}